\documentclass{article}
\usepackage{amsmath}
\usepackage{amssymb}

\usepackage[dvips]{epsfig}

\newcommand{\volltitel}[1]{}
\newcommand{\anmerkung}[1]{}

\title{Universal Sets of Quantum Gates for \\
 Detected Jump-Error Correcting Quantum Codes }
\author{G.~Alber$^1$, M.~Mussinger$^1$, A.~Delgado$^2$ \\
$^1$Institut f\"ur Angewandte Physik , Technische Universit\"at
Darmstadt, \\
D--64289 Darmstadt, Germany \\
$^2$Department of Physics and Astronomy, University of New Mexico, \\
Albuquerque,
New Mexico 87131, USA}

\begin{document}

\maketitle
\begin{abstract}
A universal set of quantum gates is constructed for the recently
developed jump-error correcting quantum codes. These quantum codes are capable
of correcting errors arising from the spontaneous decay of distinguishable
qubits into statistically independent reservoirs.
The proposed universal quantum gates are constructed with the help of Heisenberg- and
Ising-type Hamiltonians acting on these physical qubits. This way it is guaranteed that
the relevant error correcting code space is not left at any time even during the application of one
of these quantum gates. The proposed entanglement gate is particularly well suited
for scalable quantum processing units whose elementary registers are based
on four-qubit systems.
\end{abstract}


\flushbottom


\section{Introduction} 

Within the last two decades quantum information has become a vital and
fast growing research field
\cite{Bouwmeester00,Nielsen00,Alber01}. Secure key exchange
(quantum cryptography), the perfect transfer  of unknown quantum
states (teleportation) and the development of powerful quantum algorithms
\cite{Deutsch,Shor,Simon,Grover97}
demonstrate in an impressive way the pratical
potential of quantum physics. 
However, the two characteristic quantum phenomena these developments are based on, namely
interference and entanglement, are very fragile and can be destroyed easily by uncontrolled
interactions with an environment. In order to protect quantum information 
against decoherence resulting from such uncontrolled interactions,
powerful methods of quantum error correction
have been developed over the last few years. The first such code has been constructed
by Shor 
\cite{Shor95} by transfering basic ideas of error correction from the classical to the quantum domain.
This first investigation inspired
the development of various classes of active
\cite{Mabuchi96,Laflamme96,Steane96,Ekert96,Calderbank96,
Gottesman,Pellizzari96} 
and passive
\cite{Duan97,Zanardi97,Lidar98}
error correcting quantum codes.

The main aim of quantum error correction is to reverse the perturbing influence of an uncontrollable environment.
Whether such an inversion is possible or not and how it  can be achieved most efficiently
depends on the physical interaction between the quantum system considered and its environment.
In the subsequent sections we discuss main ideas underlying
a recently developed new class of error correcting quantum codes which are
capable of correcting a frequently occurring class of errors arising from spontaneous decay processes \cite{Alber01a}.
In quantum optical systems such spontaneous decay processes may arise from
the spontaneous emission of photons and in solid state devices, for example, they may originate
from the spontaneous emission of phonons. These jump codes exploit in an optimal way information
about errors which is obtained from continuous observation of the environment.
It will be demonstrated that on the basis of Heisenberg- and Ising-type
Hamiltonians universal quantum gates can be
constructed for these jump codes. They guarantee that any error can be corrected
even if it occurred during the action
of one of these gates.
Thus, with the help of these quantum gates it is possible to stabilize quantum information processing units
against spontaneous decay processes.

This contribution is organized as follows:
In Sec.~2 we summarize basic facts about the inversion of general quantum operations or generalized measurements.
One of the particularly useful results arising from the systematic analysis of this general problem is 
an algebraic criterion for the inversion of error operators.
In Sec.~3 we discuss the theoretical description of spontaneous decay processes and
of continuous measurement processes by master equations.
The practical need of inverting events involving zero- and one-photon (or phonon)  emission processes leads directly to 
one-error correcting jump codes. These quantum codes exploit in an optimal way information about
error times and error positions by monitoring the environment continuously.
In Sec.~4 we address the problem of stabilizing the coherent dynamics of a quantum system against spontaneous decay
processes. An example of such a coherent dynamics is a quantum algorithm performed by a quantum information processing
unit. In particular, we address two main problems which arise in this context. Firstly, we deal with the question
how one can implement any unitary transformation
entirely within the code space of a jump code without leaving it at any time.
Secondly, we propose a universal entanglement gate which allows one to entangle two
arbitrary basic quantum registers of a quantum information processing unit. This entanglement gate does not leave
the error correcting code space of a jump code at any time. Together with the local unitary transformations which
can be performed on any of the basic quantum registers it forms a universal set of quantum gates.

\section{Invertible quantum operations and error correction}
\label{InvertibleOp}

\subsection{Decoherence and quantum operations}
A typical quantum information processing unit is composed of a system of
$N$ two-level quantum systems, so
called qubits, which can be addressed individually.
According to the linear superposition principle of quantum mechanics
an arbitrary pure
quantum state of such a $N$-qubit quantum register is of the form
\begin{equation}
\label{DefRegister}
|\psi\rangle = \sum_{i_1,i_2,\ldots,i_N=0,1} a_{i_1 i_2 \cdots i_N} 
|i_N, \ldots, i_2, i_1 \rangle 
\end{equation}
with $|0_{\alpha}\rangle$ and $|1_{\alpha}\rangle$  denoting two
orthogonal basis states of qubit $\alpha$. 
The corresponding orthonormal basis states of the $N$-qubit Hilbert space ${\cal H}$
are denoted
$|i_N,i_{N-1},\cdots,i_1\rangle \equiv |i_N\rangle \otimes |i_{N-1}\rangle
\otimes \cdots \otimes |i_1\rangle$.
The 
complex coefficients $a_{i_1 i_2 \cdots i_N}$ fulfill the normalization condition
$\sum_{i_1,\cdots,i_N = 0,1} \mid a_{i_1 i_2 \cdots i_N}\mid ^2 = 1$. 
This generalizes easily to a system of $N$ qudits.
The power of
quantum computation relies on the ability to preserve the quantum coherence of 
such  a register-state. Any coupling to an external environment which involves uncontrollable degrees of freedom
may destroy
linear superpositions thus causing decoherence \index{decoherence} \cite{Giulini}.
This phenomenon which is undesirable from the point of view of quantum information
processing can be overcome by quantum mechanical error correction
techniques.
Shor \cite{Shor95}
demonstrated that quantum error correcting codes are possible. By now many 
different classes of error correcting quantum codes have been developed 
\cite{Mabuchi96,Laflamme96,Steane96,Ekert96,Calderbank96,Gottesman,Pellizzari96,Duan97,Zanardi97,Lidar98}. 

A main aim of quantum error correction \index{quantum error correction} is to 
reverse the dynamical influence of an external environment on the states of a quantum register
\cite{Nielsen97,Nielsen98,Caves99}.
The most general dynamical influence of this kind can be represented by a unitary joint evolution
of a quantum register with an environment followed by a von Neumann measurement performed on the
environment.
If initially the quantum register and the environment are not entangled and if the various possible
measurement results are discarded, this way a trace-preserving or \index{quantum operations!deterministic}
deterministic quantum operation
${\cal E}$ is obtained. Its action on an arbitrary register state with density operator $\rho$
(and proper normalization ${\rm Tr}\rho = 1$)
can be characterized by a set of Kraus-operators \index{Kraus-operators} \cite{Kraus}
$\{K_{lm}\}$. These Kraus- or error operators characterize
all possible environmental influences 
which may occur and they satisfy
the
completeness relation
$\sum_{lm} K^{\dagger}_{lm} K_{lm} = {\bf 1}$.
The quantum state resulting from a deterministic quantum operation is given by
\begin{eqnarray}
{\cal E}:~~\rho \to {\cal E}(\rho) &=& \sum_{l} p_l \rho_l.
\end{eqnarray}
The labels $l$ characterize all possible measurement results which occur with probabilities
$p_l = {\rm Tr}(\sum_{m} K^{\dagger}_{lm} K_{lm} \rho)$. Observation of a particular measurement result, say $l$,
 implies that immediately afterwards the register is in the normalized state
$\rho_l = \sum_{m} K_{lm} \rho K^{\dagger}_{lm}/p_l$.

Typically a set of Kraus-operators $\{K_{lm}\}$ which defines a quantum operation \index{generalized measurements}
(or generalized measurement)
is not unique. Any two sets of Kraus-operators, say $\{\overline{K}_{\lambda \mu}\}$ and $\{K_{lm}\}$,
give rise to the same quantum operation if and only if they are related by a unitary matrix ${\cal U}_{\lambda \mu, lm}$,
i.e.  $\overline{K}_{\lambda \mu} = \sum_{\lambda \mu, l m} {\cal U}_{\lambda \mu, lm}K_{lm}$ \cite{Choi}.
An important special case of deterministic quantum operations are \index{quantum operations!pure}
pure ones. They are characterized by the property that
for each measurement result $l$ the associated quantum state $\rho_l$ involves one Kraus-operator
$\{K_l \}$ only, i.e. 
\begin{eqnarray}
{\cal E}_p:~~\rho \to {\cal E}_p(\rho) &=& \sum_l p_l \rho_l
\end{eqnarray}
with $p_l = {\rm Tr}(K^{\dagger}_l K_l \rho)$, $\rho_l = K_l \rho K_l^{\dagger}/p_l$ and $\sum_l K_l^{\dagger}K_l = {\bf 1}$.
Pure quantum operations correspond to situations where a maximum amount of information about the register state is
extracted from the quantum state of an environment \cite{Nielsen97,Nielsen98,Caves99}.
As a result, an initially
prepared pure register state, say $|\psi\rangle$, remains pure, i.e.
$|\psi\rangle \to |\psi'\rangle = K_l |\psi\rangle/\sqrt{\langle \psi|K^{\dagger}_l K_l |\psi\rangle}$.

\subsection{Reversible quantum operations and error correction}
A quantum operation ${\cal E}$ is \index{quantum operations!reversible} reversible, 
if one can construct a deterministic quantum operation
${\cal R}$
such that
${\cal R}({\cal E}(\rho)) = \rho$
for any density operator $\rho$.
The recovery operation ${\cal R}$ is required to be deterministic because we want the reversal definitely to occur not
just with some probability.
In general such an inverse quantum operation cannot be constructed over the whole state space of a quantum register.
The main problem in quantum error correction
is to find an appropriate, sufficiently high dimensional subspace
${\cal C}\subset {\cal H}$
over which such an inversion operation can be defined.

It has been shown  by Knill and Laflamme \cite{Knill} and by Bennett et al. \cite{Bennett} that a quantum operation 
is reversible on a subspace ${\cal C}$ if and only if there exists a non-negative
 matrix $\Lambda_{l l'}$ such that
 \begin{eqnarray}
 P_{{\cal C}} K^{\dagger}_{l}K_{l'}P_{{\cal C}} &=& \Lambda_{l l'} P_{{\cal C}}
 \label{Knill}
 \end{eqnarray}
for all possible \index{quantum error correction!algebraic criterion} error (or Kraus-) operators $K_l$ and $K_{l'}$.
 Thereby $P_{{\cal C}}$ denotes the projection operator onto the desired subspace ${\cal C}$
 which is usually called a quantum error-correcting code space or code.
 Its code words which may be identified with classical bit-strings are formed by an orthonormal basis of states, say
 $\{|c_i\rangle, i=1,\cdots, L\}$.
The difference $r$ between the dimension of the original Hilbert space ${\cal H}$ and the dimension of ${\cal C}$, i.e.
$r = 2^N - L \geq 0$, is a measure of the \index{redundancy}
redundancy which has to be introduced in order to guarantee successful error
correction.
For the actual reversal of a quantum operation
 one has to identify first of all the character of the error (i.e.~its \index{error syndrome} syndrome) by an appropriate measurement
 and subsequently one has to apply an appropriate unitary recovery operation \index{quantum operations!recovery}
  which reverses this quantum operation \cite{Nielsen97,Nielsen98,Caves99}.
 The criterion of Eq.~(\ref{Knill}) guarantees the existence of such a measurement process and its associated unitary recovery operation. 
 These two basic steps, namely determination of the character of an error and subsequent application of a
(nontrivial) unitary recovery operation,
 constitute the basic elements of any kind of \index{quantum error correction!active} active quantum error correction.

A special situation arises, if one is able to identify a 
subspace ${\cal C}'$ which fulfills
not only Eq.~(\ref{Knill}) but also the more stringent
condition
\begin{equation}
K_{l} P_{{\cal C}'}  = \lambda _l P_{{\cal C}'} 
\label{CondDFS}
\end{equation}
for all possible error operators $K_l$  considered. In this case
 the quantity $\Lambda_{l l'}$ of Eq.~(\ref{Knill})
 factorizes according to
$\Lambda_{l l'} \equiv \lambda^*_{l}\lambda_{l'}$.
It is apparent 
that in this case
all the required unitary recovery operations are trivial as they are equal to the identity operation
over the code space ${\cal C}'$. Thus,
no recovery operation has
to be performed  at all. Such a 
passive error correction \cite{Duan97,Zanardi97,Lidar98} is \index{quantum error correction!passive} not only capable of correcting single but also multiple errors of
arbitrary order.
However, so far only very few physical situations are known in which 
sufficiently high dimensional
decoherence-free subspaces \index{decoherence-free subspace} (DFSs) ${\cal C}'$ 
can be constructed. 
In many cases the relevant DFSs
are one-dimensional so that they are not of any practical
interest for purposes of quantum information processing.

In practical applications one is interested in constructing error correcting methods
which tend to decrease not only the number of recovery operations but which also minimize redundancy.
For this purpose it may be advantageous to combine passive and active methods
of quantum error correction. 
In the subsequent sections we discuss such a family of error correcting quantum codes
which is capable of correcting spontaneous decay processes of the distinguishable 
qubits of a quantum information processor.

\section{Quantum error correction by jump codes} 
\subsection{Spontaneous decay  and quantum trajectories}
\label{QuantJumps}
\begin{center}
\begin{figure}
\centerline{
\psfig{figure=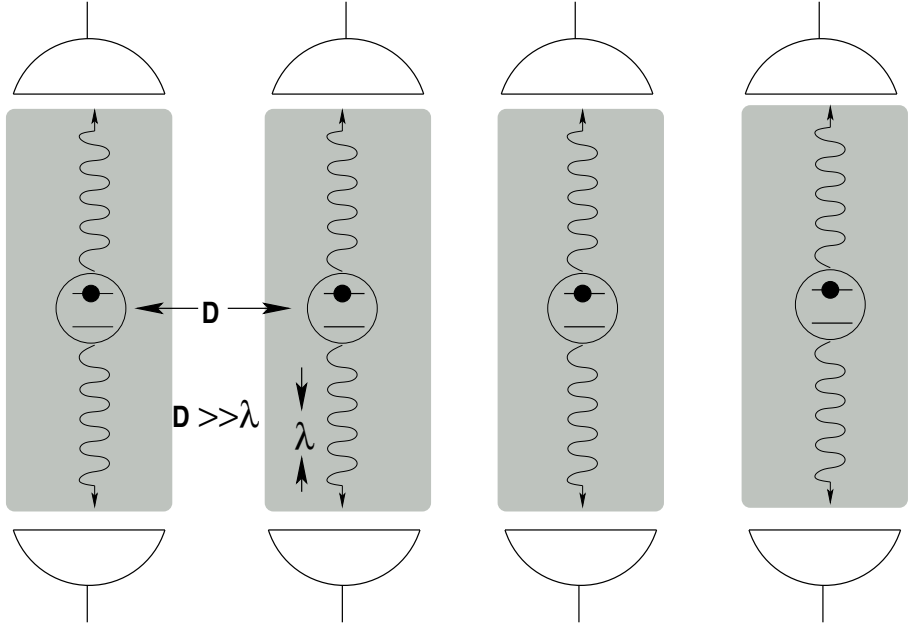,width=12cm}
}
\vspace{0.2cm}
\noindent
\caption{\label{IonTrap} Schematic representation of an array of distinguishable qubits whose spontaneous
emission times $t_i$ and error positions $\alpha_i$ are monitored continuously by photodetectors.}
\end{figure}
\end{center}

Any interaction of a quantum system with an environment whose
degrees of freedom are
not accessible to observation leads to \index{decoherence} decoherence.
An example of such an interaction is the coupling of a quantum register
to the unoccupied vacuum modes of the electromagnetic field (compare with Fig.~\ref{IonTrap}).
As a result an excited qubit can 
decay spontaneously by emission of a photon.
For the sake of quantum information processing situations are of particular interest in which
no spontaneous decay process affects the distinguishability
of the qubits involved. This is guaranteed whenever the wave lengths $\lambda$ of the spontaneously emitted
photons are much smaller than typical distances $D$  between adjacent qubits and therefore the qubits decay
into statistically independent environments.
In this case
the time evolution of the state of the quantum register 
$\rho(t)$ is given by a quantum master equation \index{quantum master equation} of the form \cite{Carmichael}
\begin{equation}
\frac{d\rho }{dt}(t)=-\frac{i}{\hbar }\left[ H, \rho(t)
\right] +\sum_{\alpha}\{\left[L_{\alpha},\rho(t) L_{\alpha}^{\dagger}\right] +
\left[L_{\alpha}\rho(t), L_{\alpha}^{\dagger}\right] \}.  \label{master equation}
\end{equation}
Thereby the 
Hamiltonian $H$ describes the coherent dynamics of 
the quantum register in
the absence of any coupling to its environment. This coherent dynamics might represent
a quantum algorithm, for example. 
The Lindblad \index{Lindblad operator} operators  \cite{Lindblad} 
$L_{\alpha} = \sqrt{\kappa_{\alpha}}|0_{\alpha}\rangle \langle 1_{\alpha}|$
with $\alpha = 1,\cdots,N$
characterize the influence of the environment
on the quantum register. The spontaneous decay rate of qubit $\alpha$ is denoted by $\kappa_{\alpha}$.
It should be mentioned that the Born- and Markov approximations underlying the derivation of
Eq.~(\ref{master equation})
are applicable whenever the interaction between system and environment is weak and, in addition, the environmental
correlation time is small. Typically these conditions are well fulfilled for quantum optical
systems. Sometimes
they are also fulfilled for other quantum systems, such as solid state
devices with phononic decay processes,
provided the environmental temperature is sufficiently high \cite{Weiss}.
\noindent

If the initial state of a quantum register is pure,
a formal solution of the quantum master
equation (\ref{master equation}) is given by \cite{Mollow,Carmichael93}
\begin{equation}
\rho \left( t\right) =\sum_{n=0}^{\infty }\sum_{{\alpha}_{1},\ldots
,{\alpha}_{n}}\int_{0}^{t}dt_{n}\int_{0}^{t_{n}}dt_{n-1}\ldots
\int_{0}^{t_{2}}dt_{1} 
 \left| t;t_{n}, \alpha_n;\ldots ;t_{1},\alpha_1\right\rangle \left\langle
t;t_{n},\alpha_n;\ldots ;t_{1},\alpha_1\right|  \label{MES1}
\end{equation}
with the unnormalized pure states
\begin{equation}
\left| t;t_{n},\alpha_n;\dots ;t_{1},\alpha_1\right\rangle = e^{-\frac{i}{%
\hbar }H_{e\!f\!f}(t-t_{n})}L_{{\alpha}_{n}}\ldots L_{{\alpha}_2} e^{-\frac{i}{\hbar }%
H_{e\!f\!f}(t_{2}-t_{1})} 
L_{{\alpha}_{1}}e^{-\frac{i}{\hbar } H_{e\!f\!f}t_{1}}\left| t=0\right\rangle .
\label{MES2}
\end{equation}
According to Eq.~(\ref{MES1})
the state of the register at time $t$ is unravelled into a sum of contributions which
are associated with all possible numbers $n$ of spontaneously emitted photons. For a
given number $n$ of emitted photons the quantum state is unravelled into a sum of
all contributions 
which describe all possible sequences of emission events 
taking place at emission times
$t_1 \leq t_2 \leq \cdots \leq t_n$
and 
affecting qubits $\alpha_1,\cdots, \alpha_n$.
The pure state
$\left|t;t_{n},{\alpha_n};\ldots ;t_{1},{\alpha_{1}}\right\rangle $
of Eq. (\ref{MES2})
describes the resulting quantum state of the register \cite{Mollow,Carmichael93}.
The quantum jumps \index{quantum jump} of the qubits from their excited to their ground states due to
spontaneous decay processes are
characterized by the Lindblad operators $L_{\alpha_i}$.  The time evolution between
two successive quantum jumps with no photon emission in between 
is described by the effective Hamiltonian 
\begin{eqnarray}
H_{e\!f\!f} &=&  H - \frac{i\hbar}{2}\sum_{\alpha} L_{\alpha}^{\dagger}L_{\alpha}.
\label{effectiveHamiltonian}
\end{eqnarray}
The norm of the quantum state of Eq.~(\ref{MES2}) yields
the probability with which a particular measurement record characterized by a
quantum \index{quantum trajectory} trajectory \cite{Carmichael93,Dalibard92,Dum} $\left(
t_{n},{\alpha_{n}};\ldots ;t_{1},{\alpha_{1}}\right) $ contributes to the density
operator $\rho \left( t\right) $. 
The formal solution of Eq.~(\ref{MES1}) describes the dynamics 
of the quantum register
under the influence of the environment in cases in which the environment is monitored continuously
by photodetectors \cite{Mollow,Carmichael93} but the measurement results are discarded.
According to Sec.~\ref{InvertibleOp}
the formal solution of Eq.~(\ref{MES1}) describes a deterministic quantum process where each quantum
trajectory characterizes a particular measurement record.

\subsection{Detected jump-error correcting quantum codes \index{jump code} }
\label{TheCode}

How can we stabilize a quantum system, such as the one depicted in Fig.~\ref{IonTrap}, against spontaneous
decay processes, if we are able to monitor the distinguishable qubits continuously by photodetectors?
According to Eq.~(\ref{MES1}) we have to tackle two major tasks.
Firstly, we have to correct the modifications taking place during successive photon emission
events. These modifications are described by the effective (non-hermitian) Hamiltonian
of Eq.~(\ref{effectiveHamiltonian}).
Secondly, we have to invert each quantum jump which is caused by the spontaneous emission
of a photon. These quantum jumps are described by the Lindblad operators
appearing in Eq.~(\ref{MES2}).

For the sake of simplicity let us concentrate in this section on the case of a quantum memory
without any intrinsic coherent time evolution, i.e. $H\equiv 0$ in Eq.~(\ref{master equation}).
If we want to correct the errors taking place during successive photon emission events, we
must
invert the pure quantum operation which is characterized by the one-parameter family of Kraus-operators
\begin{eqnarray}
K_0(t) &=& e^{- \sum_{\alpha}L^{\dagger}_{\alpha}L_{\alpha} t/2}.
\end{eqnarray}
Specializing 
the criterion of Eq.~(\ref{Knill}) 
to the case of these hermitian error operators an inversion
is possible over a subspace ${\cal C}$ if and only if
\begin{eqnarray}
P_{{\cal C}} K_0(2t) P_{{\cal C}} = \Lambda_{00}(t) P_{{\cal C}} 
\end{eqnarray}
with $\Lambda_{00}(t) \geq 0$. Stated differently, over the code space ${\cal C}$
the undesired modification appearing 
in the effective Hamiltonian of Eq.~(\ref{effectiveHamiltonian}) has to act as a (non-negative) 
multiple of the unit operator. Thus, the code space we are looking
for is a DFS of the effective Hamiltonian with $H\equiv 0$.

In the subsequent discussion we focus on
the important special case in which the spontaneous decay rates of all qubits are equal,
i.e. $\kappa_{\alpha} \equiv \kappa$. The corresponding DFSs can be found easily
because the relevant operator, 
i.e.~$\sum_{\alpha}L^{\dagger}_{\alpha}L_{\alpha} = \kappa \sum_{\alpha} |1_{\alpha}\rangle
\langle_\alpha 1|$, just enumerates the number of excited qubits.
Therefore, any set of orthonormal states which all involve the same number of excited qubits
constitutes a passive error correcting code space for the Kraus-operators $K_0 (t)$. 
The dimension $D$ of a DFS
 involving $N$ physical qubits $k$ of which are excited, i.e. a ${\rm DFS}-(N,k)$,
is given by 
\begin{equation}
D = {N \choose k} \equiv \frac{N!}{k! (N-k)!}.
\end{equation}
For a given number of
physical qubits $N$ this dimension is maximal, if half of the qubits are excited, i.e.~for $k = [N/2]$.
($[x]$ denotes the largest integer smaller or equal to $x$.)
Such a DFS
of maximal dimension involving four physical qubits, for example,
is formed by the set of code words $\{|1100\rangle,$ $|0011\rangle,$ 
$|1010\rangle,$ $|0101\rangle,$ 
$|1001\rangle,$ $|0110\rangle \}$.

In general, arbitrary linear superpositions of code words 
of such a DFS cannot be stabilized against quantum jumps arising from spontaneous decay processes.
If we also want to invert each individual quantum jump, we have to find an appropriate subspace
${\cal C'}\subseteq {\cal C}$ over which any of the quantum operations appearing in Eq.~(\ref{MES1})
is reversible. For this purpose we note, that
within any ${\rm DFS}-(N,k)$
the time evolution between successive quantum jumps
is proportional to the unit
operator, i.e.~$e^{- \sum_{\alpha}L^{\dagger}_{\alpha} L_{\alpha} t/2}|_{{\cal C}} \equiv
e^{- k \kappa t/2} P_{{\cal C}}$. Therefore, we have to find appropriate subspaces ${\cal C'} \subseteq {\cal C}$
over which the Lindblad operators appearing in Eq.~(\ref{MES2}) are reversible.
The details of the
construction of an active error correcting quantum code capable of correcting one quantum jump at a time, for
example, depends very much on whether 
the error position is known or not.
In the case of an unknown error position one has to fulfill the criterion of Eq.~(\ref{Knill}) for
all possible Lindblad operators $L_{\alpha}$ with $\alpha \in \{1,\cdots,N\}$.
Plenio et al.~\cite{Plenio} have been able to find such a code which requires at least
eight physical qubits for the  encoding of one logical qubit, i.e.~for two orthonormal logical
states. In contrast, if the error position $\alpha$ of a quantum jump characterized by Lindblad
operator $L_{\alpha}$ is known, the  redundancy \index{redundancy} of such an active one-error correcting
quantum code which is embedded into a passive code can be lowered significantly.

The simplest example of such an embedded quantum code or jump code 
\index{jump code} \index{embedded quantum code}
which is capable of correcting one error at a time
can be constructed with the help of four physical qubits \cite{Alber01a}. The (unnormalized) code words of this
particular jump code represent a logical
qutrit and are given by
\begin{eqnarray} \nonumber 
|c_{0}\rangle &=&|0011\rangle + e^{i \varphi} |1100\rangle ,  \\
\label{423} |c_{1}\rangle  &=& |0101\rangle + e^{i \varphi} |1010\rangle ,\\
\nonumber
|c_{2}\rangle  &=& |0110\rangle + e^{i \varphi} |1001\rangle  
\end{eqnarray}
with an arbitrary phase $\varphi$.  
Obviously, the code words of this jump code consist of four-qubit states in which half of
the qubits are excited.
The equal number of excited qubits 
involved in this code
guarantees that the effective time evolution between successive
quantum jumps 
is corrected passively.
A characteristic feature of this quantum code is the complementary pairing of states with
equal probabilities.
This latter property guarantees the validity of the necessary and sufficient
conditions of Eq.~(\ref{Knill}) provided the error position is known. 
This one-error correcting jump code 
involves three logical states and four physical qubits two of which are excited. Therefore,
let us call it jump code $1-JC(4,2,3)$.
This construction of a one-error correcting embedded quantum code
can be generalized easily to any  even number $N$ of physical qubits.
Thus, any jump code $1-JC(N,N/2,{N-1\choose N/2 -1})$  can be constructed 
by an analogous
complementary paring of $N$-qubit states half of which are excited.
This way one obtains ${N-1\choose N/2-1}$ orthogonal code words which
form a one-error correcting embedded quantum code for spontaneous decay processes.
It can be shown that this family of one-error correcting quantum codes is optimal 
in the sense that their redundancy cannot be reduced any further \cite{Alber01a}.
Asymptotically, for large numbers of physical qubits the effective number of logical qubits $L_q$
which can be encoded by the jump code $1-JC((N,N/2,{N-1\choose N/2 -1})$ is given by
$
L_q \equiv {\rm log}_2 {N-1\choose N/2 -1} = N - {\rm log}_2\sqrt{N} + O(1).
$
In addition,
 far reaching links between these jump codes and fundamental structures of combinatorial design
theory \cite{Beth99} can be established. These links
are expected to be particularly useful for the further development of many-error correcting
embedded quantum codes with low redundancy.

\begin{figure}
\centerline{\psfig{figure=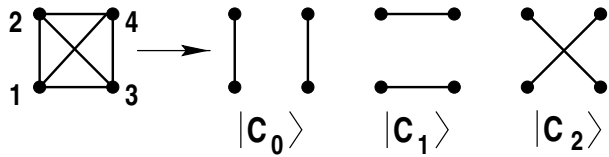,width=8.0cm,clip=}}
\caption{Affine plane of four points over the binary field and its associated parallelisms.}
\label{Fig1}
\end{figure}

In order to demonstrate some basic aspects of these links let
us consider the previously discussed optimal $1-JC(4,2,3)$-code as an example.
This embedded quantum code is constructed within the six-dimensional DFS 
which involves all quantum states of four qubits two of which are excited.
These six quantum states can be represented graphically by six
lines as depicted in Fig.~\ref{Fig1} on the left hand side.
Each point in this diagram represents a qubit.  Each basis state
of this DFS is represented by a line connecting the two qubits which are
excited.
\noindent
This system of points and lines has a few interesting properties, namely
\\
(1)~~any two points define a unique line;\\
(2)~~there are at least two points on each line;\\
(3)~~there are three points which are not on a line;\\
(4)~~to each line $g$ and each point $P$ not contained in $g$ 
there exists a uniquely \\ \hspace*{0.5cm} determined
parallel
line $h$ which has no point in common with $g$
(axiom of\\\hspace*{0.5cm}
 parallels).\\
In combinatorial design theory  a structure fulfilling these axioms is called an affine plane.
The three code words of our previously discussed $1-JC(4,2,3)$-code 
(compare with Eq.(\ref{423}))
correspond to the three possible parallel pairs of this affine plane.
Thus, the affine plane of Fig.~\ref{Fig1} may be viewed as a generating design for the parallelisms 
which are associated with the basis states of the 
$1-JC(4,2,3)$-code. Exploiting this link jump codes can  be constructed which 
are even capable of 
correcting more than one error at a  time \cite{Alber01a,Beth2002}.

Provided the decay rates of all qubits are equal, error position and error time can
be determined perfectly and
recovery operations are applied immediately after the observation of a quantum jump,
spontaneous decay processes can be corrected perfectly with these jump codes.
But in reality, typically none of these conditions is
fulfilled precisely. However, numerical simulations demonstrate that quantum states
can be stabilized against various types of imperfections still to a high degree even if
some of these conditions are not fulfilled perfectly \cite{Alber02b}.

\section{Universal sets of quantum gates for detected jump-error correcting code
spaces } 

The previously discussed error correcting jump codes allow one to
stabilize a quantum memory against spontaneous decay processes.
However, in order to be useful also for purposes of quantum information processing and quantum
computation 
two major additional requirements have to be fulfilled.
Firstly, one should be able to manipulate pure quantum states 
in such a way that a chosen error correcting code space
is not left at any time during the performance of a quantum algorithm.
This can be achieved by using a universal set of quantum gates which operates
entirely within  an error correcting code space and which is
implemented by a set of 
Hamiltonians leaving this
code space invariant. Such a Hamiltonian implementation of universal
quantum gates guarantees that
any quantum algorithm which is implemented with the help of these
quantum gates does not leave this code space at any time even during the application
of one of these quantum gates.
Secondly, analogous to classical computer architecture,
it is desirable
to develop quantum information processors
which are based on small quantum registers and, in addition, to design quantum gates
in such a way that in each step at most two basic quantum registers are entangled. This ensures
that the same set of quantum gates can be used for an arbitrarily large quantum information processing
unit. For a recent proposal on implementing these ideas on suitable subspaces of our detected
jump-error correcting codes see \cite{Khodjasteh}.
In the subsequent sections we present an example
how a quantum information processing unit
meeting these two major requirements can be constructed
on the basis of elementary
four-qubit registers each of which constitutes
a local qutrit of the jump code $1-JC(4,2,3)$.



%

\subsection{Universal sets of quantum gates for qudit-systems}
Universal sets of quantum gates \index{universal quantum gates} for qubit-systems were considered by D.~DiVincenzo
\cite{Vincenzo95a}, A.~Barenco et al.~\cite{Barenco95a} and S.~Lloyd
\cite{Lloyd95a}. These authors have shown that with a few Hamiltonians acting on single
qubits and with one particular two-qubit Hamiltonian it is possible to generate any
unitary transformation for a quantum register consisting of qubits.
All possible one-qubit operations 
are members of the continuous group $SU(2)$ (suppressing a trivial $U(1)$ operation) and the two
qubit operation entangles any two separable qubits.
The lowest dimensional 
member of our previously discussed jump codes, namely the $1-JC(4,2,3)$-code, provides 
a logical qutrit and therefore the most general unitary qutrit-operations
needed for quantum information processing within this code space
are members of the continuous group $SU(3)$
(again suppressing a trivial $U(1)$ operation).
Thus the natural
question arises which set of  quantum gates is universal and thus 
capable of generating an arbitrary unitary transformation within the state space of
a qutrit. 

Jean-Luc and Ranee Brylinski \cite{Brylinski01} derived  a generalization of the
results of D.~DiVincenzo, A.~Barenco et al.~and S.~Lloyd.  In particular, they
demonstrated that for $d$-dimensional elementary data carriers, so called qudits,
every N-qudit gate
can be  obtained by combinations of all one-qudit gates and a certain two-qudit
entanglement gate.
In particular, these authors call a collection ${\cal G}$
of one-qudit and two-qudit gates 
universal (exactly universal), if every $N-$qudit gate with $N \geq 2$ can be
approximated with arbitrary accuracy (represented exactly)
by a circuit made up of $N$-qudit  gates of this collection 
${\cal G}$. A (unitary) two-qudit gate $V$ is called primitive, if it
maps separable pure states again to separable pure states. Thus, if $|x\rangle$ and $|y\rangle$ are qudit-states,
we can find qudit-states
$|u\rangle$ and $|v\rangle$ such that $V |x\rangle|y\rangle = |u\rangle|v\rangle$. If $V$ is not
primitive, it is called  imprimitive. 
Suppose we are given a two-qudit gate $V$. Then 
the collection of all one-qudit gates together with $V$ is universal
if and only if $V$ is imprimitive.
In particular, J.-L. and R. Brylinski \cite{Brylinski01} have proved the useful criterion \index{universal quantum gates!criterion} that,
if a (unitary) two-qudit gate $V$ is diagonal in a
computational basis, i.e. $ V |j\rangle|k \rangle = \exp ( i \theta_{jk} ) |j\rangle|k \rangle $,
$V$ is primitive if and only if we have  
\begin{equation}
 \theta_{jk} + \theta_{pq}  \equiv \theta_{jq} + \theta_{pk}~~ ( {\rm mod} 2 \pi ) 
 \label{DiagonalEntanglingGate}
\end{equation}
for all possible values of $j,k,p,q$.

In general,
the difficulty of finding an appropriate set
of Hamiltonians by which one can generate a universal set of quantum gates operating
entirely within an error correcting code space depends on the physical interactions available.
Typical physical two-body interaction Hamiltonians
which are expected to be realizable in laboratory
are Heisenberg and Ising Hamiltonians $H_{He}$ and $H_{Is}$, i.e.
\begin{eqnarray} \nonumber
H_{He} & = & \sum_{\alpha \beta} C_{\alpha \beta}(t)
(\sigma_{\alpha}^{(x)} \sigma_{\beta}^{(x)}+\sigma_{\alpha}^{(y)}
\sigma_{\beta}^{(y)}+\sigma_{\alpha}^{(z)} \sigma_{\beta}^{(z)}), \\
 H_{Is} & = & \sum_{\alpha \beta} D_{\alpha \beta}(t)
\sigma_{\alpha}^{(z)} \sigma_{\beta}^{(z)}.
\label{HIsing}
\end{eqnarray}
Thereby, $\sigma_{\alpha}^{(x)},\sigma_{\alpha}^{(y)},\sigma_{\alpha}^{(z)}$
denote the three Cartesian components of the Pauli spin operators of qubit $\alpha$ and
the quantities $C_{\alpha \beta}(t)$ and
$D_{\alpha \beta}(t)$ denote coupling coefficients of qubits $\alpha$ and $\beta$.
These latter coefficients are assumed to be tunable arbitrarily.
If it is not possible to realize particular linear combinations or commutators
of these
Hamiltonians by appropriate tunings of these coupling coefficients, one may
use appropriate products, such as
\begin{eqnarray}
 e^{i( t_1 H_1 + t_2 H_2) } &=& \left( e^{i \frac{t_1}{n} H_1 } e^{i
\frac{t_2}{n} H_2
}\right) ^n + O\left( \frac{1}{n} \right),
\nonumber  \\ \label{ProdApprox2}
 e^{i( i[ t_1 H_1,  t_2 H_2 ]) } &= &\left( e^{i \frac{t_1}{\sqrt{n}} H_1 }
e^{i \frac{t_2}{\sqrt{n}} H_2} e^{-i \frac{t_1}{\sqrt{n}} H_1 } e^{-i
\frac{t_2}{\sqrt{n}}
H_2 }\right) ^n + O\left( \frac{1}{\sqrt{n}} \right).
\end{eqnarray}
According to Eqs.(\ref{ProdApprox2}) one needs infinite products for representing unitary
transformations corresponding to sums or commutators of Hamiltonians exactly.
However, it can be shown that in many cases exact representations
can also be obtained which involve finite products only \cite{DAlessandro01,Zeier02}.

\subsection{Universal one-qutrit gates}
In this section we address the question how arbitrary unitary transformations can be
implemented in the error correcting code spaces of jump codes with the help of Heisenberg-type
and Ising-type Hamiltonians. Thereby the Hamiltonians considered are expected to leave these code spaces
invariant so that during the application of an arbitrary sequence of unitary transformations
the error correcting code space is not left at any time. This requirement guarantees that any error
due to a spontaneous decay process occurring during the processing of a quantum state
can be corrected. As an example we consider the implementation of arbitrary unitary
transformations in the lowest dimensional one-error correcting
jump code, i.e. the $1-JC(4,2,3)$-code \cite{Alber01c}.

Two classes of two-particle
Hamiltonians of the Heisenberg- and Ising-type acting on
physical qubits will be needed for this construction, namely
\begin{eqnarray}\nonumber
E_{\alpha \beta} &  =  & \frac{1}{2}  \left(
P_{\alpha,\beta}  + \sigma_{\alpha}^{(x)} \sigma_{\beta}^{(x)}
 + \sigma_{\alpha}^{(y)} \sigma_{\beta}^{(y)}  +
 \sigma_{\alpha}^{(z)} \sigma_{\beta}^{(z)}
\right), \\
F_{\alpha \beta} & = & \frac{1}{2}   \left(
P_{\alpha,\beta} + \sigma_{\alpha}^{(z)} \sigma_{\beta}^{(z)} \right)
\label{PhaseHam}
\end{eqnarray}
with $\alpha, \beta = 1,\cdots, N$.
Any member of this family of
two-particle Hamiltonians acts on the physical qubits
$\alpha$ and $\beta$ only leaving all other qubits unaffected.
The terms of Eqs.(\ref{PhaseHam})
involving the  projection operator $P_{\alpha, \beta}$
represent an energy shift of the two qubits.
The residual interaction terms are of Heisenberg- and Ising-type.
>From these Hamiltonians we can select the six members
$E_{12}$, $E_{23}$, $E_{13}$, $F_{12}$, $F_{13}$, $F_{13}$, for example.
\begin{table}
\begin{center}
\begin{tabular}{l|llllll} \hline
$E_{\alpha\beta}$ & $E_{12} $ & $E_{23}$  & $E_{13}$ & $F_{12}$  
& $F_{23}$  & $F_{13}$ \\ \hline 
 $|c_0\rangle $ \rule{0mm}{3ex} & $|c_0\rangle $ & $|c_1\rangle $ & 
$|c_2\rangle $ & $|c_0\rangle $ & 0 & 0 \\
$|c_1\rangle $ & $|c_2\rangle $ & $|c_0\rangle $ & 
$|c_1\rangle $ & 0 & $|c_1\rangle $  & 0 \\
$|c_2\rangle $ & $|c_1\rangle $ & $|c_2\rangle $ & 
$|c_0\rangle $ & 0 &  0 & $|c_2\rangle $   \\ \hline
\end{tabular}
\caption{\label{OperationTab} 
Action of the Hamiltonians $E_{12}, E_{23}, E_{13}$
and $F_{12},F_{23}, F_{13}$ on the
code words of a detected jump-error correcting quantum code consisiting of
four qubits with a phase $\varphi =0$ (see Eq.(\ref{423})).}
\end{center}
\end{table}
Their action on the code words of the jump code $1-JC(4,2,3)$  with
$\varphi =0$ (compare with Eq.~(\ref{423})) is  given in Table 
\ref{OperationTab}  and can be represented
by the matrices
\begin{eqnarray}
E_{12} = \left(
\begin{array}{rrr}
1 & 0 & 0 \\
0 & 0 & 1 \\
0 & 1 & 0
\end{array} \right),
  &
 E_{23} = \left(
\begin{array}{rrr}
0 & 1 & 0 \\
1 & 0 & 0 \\
0 & 0 & 1 
\end{array} \right),
 &
E_{13} = \left(
\begin{array}{rrr}
0 & 0 & 1 \\
0 & 1 & 0 \\
1 & 0 & 0 
\end{array} \right), \\  \nonumber
F_{12} = \left( 
\begin{array}{rrr}
1 & 0 & 0 \\
0 & 0 & 0 \\
0 & 0 & 0 
\end{array} \right),
&
F_{13} = \left( 
\begin{array}{rrr}
0 & 0 & 0 \\
0 & 1 & 0 \\
0 & 0 & 0 
\end{array} \right),
 & 
F_{23} = \left( 
\begin{array}{rrr}
0 & 0 & 0 \\
0 & 0 & 0 \\
0 & 0 & 1
\end{array} \right).
\end{eqnarray} 
Accordingly, the unitary transformations resulting from the Hamiltonians 
$E_{12}, E_{13}$ and
$E_{23}$ swap two codewords and change the phase of the third code word. The unitary
transformations
resulting from the Hamiltonians $F_{12}, F_{13}$ and $F_{23}$ change the phase of exactly one
of the three 
code words.
It is straight forward to demonstrate that the
six operators
\begin{eqnarray}
C_{12}^+ = E_{23} - F_{23},~ &~ C_{13}^+= E_{13} - F_{13},~&~
C_{23}^+ = E_{12} - F_{12}, \\ \nonumber
 C_{12}^- = i[C_{13}^+, C_{23}^+],~&~
C_{13}^- = i[C_{12}^+, C_{23}^+],~& ~ C_{23}^- = i[C_{12}^+, C_{13}^+]  
\end{eqnarray}
and the two operators $F_{12}$ and $F_{13}$ form a basis for the Lie Algebra of
the continuous group 
$SU(3)$. 
Thus, by an appropriate linear combination of these eight generators 
any unitary transformation belonging to the continuous group $SU(3)$
can be represented
on the code space of the jump code $1-JC(4,2,3)$.

\subsection{A universal entanglement gate} 
\noindent
\begin{center}
\begin{figure}
\centerline{
\psfig{figure=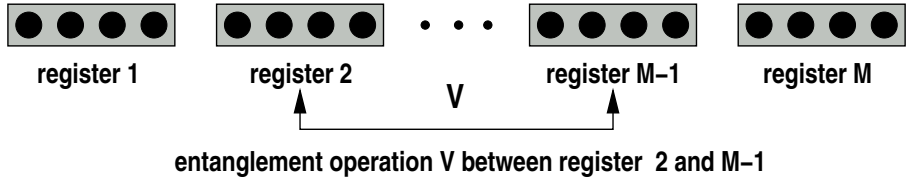,width=12cm}
}
\vspace{0.2cm}
\noindent
\caption{\label{register} Schematic representation of an array of qubits
consisting of $M$ basic registers. Each basic register carries a logical qutrit of the jump code
$1-JC(4,2,3)$. 
Any two basic registers, such as registers 2 and $(M-1)$,
can be entangled by the entanglement  gate $V$.}
\end{figure}
\end{center}
In computer science it is common practice to use  basic registers \index{registers} of a fixed size
and to scale an information processing unit by using several of these basic registers.
Consequently, on the one hand
an algorithm consists of the
manipulation of single basic registers, and on the other hand of the interaction
between any two of these registers at a time.
Such an architecture ensures that the same set of
gates can be used for an arbitrarily scaled device.
In addition, new registers can be added to the information processing unit at  any
time even during a computation without necessitating a new encoding of all
qubits involved.
If one applies this idea to a quantum processing unit, the basic registers
are formed by an appropriate number of qubits. In addition, if one wants
to correct errors originating from spontaneous decay processes, the simplest
basic register has to consist of four physical qubits which form a jump code
$1-JC(4,2,3)$.
Thus, an appropriate quantum information processing unit capable of stabilizing quantum
algorithms against spontaneous decay processes
would consist of an array of such four-qubit clusters (compare with Fig.~\ref{register}).
We have already demonstrated in the
previous section that any unitary transformation within such a four-qubit basic quantum register
can be implemented with the help of Heisenberg- and Ising-type Hamiltonians.
Here we present a universal entanglement gate which is capable of entangling
two arbitrary four-qubit basic registers and which is based on Ising-type Hamiltonians.
Together with the unitary transformations discussed in the previous section this entanglement gate
forms a universal
set of quantum gates for a quantum information processing unit which is based on four-qubit registers.
In addition,
the presented
entanglement gate ensures that all errors due to spontaneous decay processes can be
corrected even if they take place during 
the application of a quantum gate.


Let us consider first of all  the nine tensor product states which are associated with
two  basic four-qubit registers. These states are constituted by the product states of two jump codes $1-JC(4,2,3)$, namely 
\begin{eqnarray*}
|00\rangle_L &=&  
|00110011\rangle +  |11001100\rangle \; + \; |00111100\rangle +
|11000011\rangle,  \\
|01\rangle_L  &= &
 |00110101\rangle + |11001010\rangle  \; + \; |00111010\rangle +
|11000101\rangle,  \\
|02\rangle_L &=& 
|00110110\rangle + |11001001\rangle \;  +  \; |00111001\rangle +
|11000110\rangle,  \\
|10\rangle_L &=& 
|01010011\rangle + |10101100\rangle \;  +  \; |01011100\rangle +
|10100011\rangle,  \\
|11\rangle_L &=&
|01010101\rangle + |10101010\rangle \;  +  \; |01011010\rangle +
|10100101\rangle,  \\
|12\rangle_L &=& 
|01010110\rangle + |10101001\rangle \;  + \;  |01011001\rangle +
|10100110\rangle, \\
|20\rangle_L &=& 
|01100011\rangle + |10011100\rangle \; +  \; |01101100\rangle +
|10010011\rangle, \\
|21\rangle_L &=& 
|01100101\rangle + |10011010\rangle \; + \; |01101010\rangle  +  
|10010101\rangle,  \\
|22\rangle_L &=& 
|01100110\rangle + |10011001\rangle \; + \; |01101001\rangle +
|10010110\rangle. 
\end{eqnarray*}
The linear subspace spanned by these states is denoted by
${\cal C}_9$.
It is apparent that these states are
linear superpositions of code words of the one-error correcting 
jump code $1-JC(8,4,35)$. 
Let us assume that it is possible to implement the Ising-type Hamiltonian
\begin{eqnarray}
H_{ent} &=& 1/2 ( F_{26} + F_{36} + F_{27} + F_{37})
\end{eqnarray}
by an appropriate tuning of the coupling coefficients of Eq.~(\ref{HIsing}).
This Hamiltonian leaves the code space of the jump code $1-JC(8,4,35)$
invariant so that any spontaneous decay process can be corrected. 
Let us denote
the linear subspace spanned by the eight orthonormal states
\begin{equation}
\{|00\rangle_L,|01\rangle_L,|02\rangle_L,|10\rangle_L,|11\rangle_L,|12\rangle_L,|20\rangle_L,
|21\rangle_L\}
\end{equation}
 by $A$ and the subspace spanned by the two orthonormal states
\begin{equation} |22+\rangle_L =|01100110\rangle + |10011001\rangle
\end{equation}
 and 
\begin{equation}
|22-\rangle_L =|01101001\rangle + |10010110\rangle 
\end{equation} 
 by B.
With this notation the action of the Hamiltonian 
can be represented by
$
H_{ent} = P_A \oplus 2|22+\rangle_{LL}\langle 22+|  
$
with $P_A$ denoting the projection operator onto subspace $A$.
Therefore, the Hamiltonian $H_{ent}$ acts in the subspaces $A$ and $B$ differently. 
Applying this Hamiltonian for the (dimensionless) time $\tau$ yields the unitary transformation
 \begin{equation}
 U(t) =  e^{-i H \tau } = e^{-i \tau} P_A \oplus(e^{-i2\tau}|22+\rangle_{LL}\langle 22+|
 + |22-\rangle_{LL}\langle 22-|).
\label{Ut}
 \end{equation} 
Though states $|22+\rangle_L$ and $|22-\rangle_L$ are affected differently by this Hamiltonian
the unitary transformation of Eq.~(\ref{Ut}) does not leave the one-error
correcting code space $1-JC(8,4,35)$
at any time.
Therefore, any spontaneous emission event can be corrected.         
In order to implement an entanglement operation within the tensor product space of two
basic four-qubit registers
we choose the (dimensionless) interaction time so that
$\tau = \pi$. This implies that 
all code words in subspace $A$ are multiplied by a factor $(-1)$ and states
$|22+\rangle_L$ and $|22-\rangle_L$   are both multiplied by  a factor $(+1)$.
Applying
an additional global factor of $(-1)$ results in the conditional phase gate $V$ 
 \begin{equation} 
 V  = P_A  - |22\rangle_{LL}\langle 22|.
\end{equation}
This conditional phase gate is a universal entanglement gate because,
consistent with the notation of Eq.~(\ref{DiagonalEntanglingGate}),
$\theta_{ij} = 0$ for all $(i,j) \neq (2,2)$ and  $\theta_{22} = \pi$.
Therefore, 
$ \theta_{12} + \theta_{21} = 0 \not\equiv \pi = \theta_{11} + \theta_{22}
~~({\rm mod} 2 \pi ) 
$
and according to the criterion of Eq.~(\ref{DiagonalEntanglingGate}) $V$ is a universal entanglement gate.

\section{Summary and outlook} 

We discussed main ideas underlying a recently introduced class of error correcting quantum codes,
the so called jump codes,
which are capable of correcting spontaneous decay processes originating from the coupling of distinguishable
qubits to statistically independent environments. These quantum codes exploit
information about error times and error positions 
in an optimal way by monitoring the environment continuously. 
We also addressed the practical question how these error correcting quantum codes can be used for 
stabilizing a quantum algorithm against these types of errors. For this purpose we presented
a set of universal quantum gates which guarantees that any error due to a spontaneous 
decay process can be corrected even if it occurred during the application of one of these quantum gates.
This is possible because these quantum gates are based on Heisenberg- and Ising-type Hamiltonians which leave
the code space of a jump code invariant. 

Though our discussion concentrated on one-error correcting quantum jump codes, the already mentioned connection
with basic concepts of combinatorial design theory may offer interesting perspectives also for the construction 
of multiple-error correcting jump codes with minimal redundancy. Such optimal multiple-error correcting
quantum codes are expected to be particularly useful for stabilizing the dynamics of quantum information processing
units against environmental influences.

\section{Acknowledgments}
This work is supported by the Deutsche Forschungsgemeinschaft. Discussions with T.~Beth, I.~Cirac,
M.~Grassl, R.~Laflamme, D.~Lidar and  D.~Shepelyansky are gratefully acknowledged.

\end{document}